# Functional Nanomaterials Design in the Workflow of Building Machine-Learning Models

Zhexu Xi

*Bristol Centre for Functional Nanomaterials, University of Bristol, Bristol, UK*

**Abstract:** Machine-learning (ML) techniques have revolutionized a host of research fields of chemical and materials science with accelerated, high-efficiency discoveries in design, synthesis, manufacturing, characterization and application of novel functional materials, especially at the nanometre scale. The reason is the time efficiency, prediction accuracy and good generalization abilities, which gradually replaces the traditional experimental or computational work. With enormous potentiality to tackle more real-world problems, ML provides a more comprehensive insight into combinations with molecules/materials under the fundamental procedures for constructing ML models, like predicting properties or functionalities from given parameters, nanoarchitecture design and generating specific models for other purposes. The key to the advances in nanomaterials discovery is how input fingerprints and output values can be linked quantitatively. Finally, some great opportunities and technical challenges are concluded in this fantastic field.

## 1. Introduction

In the research field of nanomaterials, discovering novel materials with targeted functionalities and more explicit structure-property correlations has always been a hotspot. Assorted traditional methods like trial-and-error methods have been widely put into practice to improve the efficiency and accuracy of developing new materials[1][2]. However, endless, insurmountable shortcomings, such as experimental factors (e.g. lengthy and tedious experimental periods, extreme experimental conditions and insufficient operation experience) and theoretical limitations (e.g. microstructures, surface and bulk states and different electron motion), severely restricts the rapid progress of materials. In addition, versatile characterizing techniques bring about the complexity of real-world materials based on multi-forms and multi-levels of massive databases[3-6]. Moreover, the emergence of advanced computational-chemistry-based methods, like Monto Carlo tools, First Principle calculations, and Molecular Mechanics, contributes to the specific structure-property analysis for limited materials, not scaling to sophisticated systems and practical problems[7]. Accordingly, the ML-assisted approaches have been emphasized in pace with the combination of artificial intelligence (AI) and big data (usually known as the fourth scientific research paradigm).

ML, a branch of AI, can effectively derive useful information from known and hypothesized systems by training a set of big material data, sometimes surpassing the human-expert capabilities and reaching the unprecedented particulate scale. ML-assisted techniques here can make accurate predictions of material properties and behaviors from existing data and then generalize more data for model refinement[8][9]. Simultaneously, high performance efficiency and time efficiency are guaranteed when tackling complex molecular or material systems[10].

In this article, we give a comprehensive, in-depth summary of the typical features of ML in nanomaterials discovery, design and further large-scale application in the workflow of constructing ML models: identifying the targets, collecting the available data, feature engineering, model selection and training, and application.

The underlying structure-property correlations are presented for future trends and perspectives in the ML-driven nanomaterials discovery and design process. We aim to broaden the horizons of readers for the ML-based tools in further nanomaterials-related research.

## 2. Complete process of machine learning in nanomaterials design and discoveries

The attractive ability of machine learning in materials science is to generalize a set of unknown data and obtain form-fitting physical laws by training data without human input for more accurate output. This advanced approach provides a new insight into various state-of-the-art applications in nanomaterials, like high-efficiency energy conversion and storage[10][11], electron transfer behavior in nanoparticles[12], and optimized molecular recognition[13].

Based on the above five critically basic steps in the workflow of constructing ML models, the complete nanomaterials-related ML process can be categorized into two classes: feature engineering aiming from raw database to features (mainly consisting of human-induced engineering previously and more prevalent automated one in future), and modelling aiming from features to models. Specifically, several descriptors, which correspond to a series of programmable structural or property-related parameters, are first introduced to stand for the raw nanomaterials; second, these characteristic parameters are input for data training and then giving access to ML model construction; third, this generated model needs to be extended to an applicable level to predict new properties at nanoscale and determine novel nanomaterials with new nanosurfaces or nanostructures, including quantitative composition-structure-function relationship and structure-activity relationship[2-4,13-16].

## 3. Identification of purposes and targets

The first step of ML model construction is to identify the prediction purposes and targets, which is primarily determined by the knowledge supplement and research field of professions or experts. The suitable purposes potentially give access to a durable and high-performance model; otherwise, the wrong choice of targets may bring about a fake error or uncertain accident. For example, in ML models, the formation energy at 0 K $E_f$ or the energy above the convex hull $E_{hull}$ is used to describe the thermodynamic stability in the design of nanopatterns[17,18]. The difficulty in serving $E_{hull}$ as a target function originates from the multi-phase competition in the phase diagram; $E_f$ seems a properer choice, but it also counts on the choice of the reference states. Specifically, the value of the elemental ground states, which are usually be reckoned as the reference states, is closely linked with the DFT-based errors, especially entailing redox reactions[19]. Furthermore, the $E_{hull}$ value can be derived from the $E_f$, and then its stability can be precisely classified with the application of the appropriate threshold (e.g. a threshold of 0 indicating the phases on the hull; a positive threshold explains the metastability in ML calculations)[20,21].

In addition, aiming at what kinds of materials problems also makes a difference in ML models selection, like whether the property parameters are categorical or regression-based. The difference is the nature of the mapping: from input to categorical targets in classification analysis and from input to target values in regression analysis. Note that regulating the proper value of threshold can realize the conversion between both kinds of models[22].

## 4. Data collection

Collecting training data is vital to identify and address the missing or spurious elements and improve the accuracy, speed and reputability of ML algorithms. The sources of data can be publicly provided or generated via experiments or computations. Specifically, the available sources of data with various qualities and quantities can be divided into four parts: experimental or simulated materials properties (e.g. physical, chemical, structural, thermodynamic and kinetic properties); chemical-reaction-based data (e.g. reaction rates, temperature and pressure); the existing data from the literature; graphical data.

The high-quality and abundant-quantity data is critically concerned in many peer-reviewed research fields, like cheminformatics. In this field, many experts with enriched domain knowledge obey these guidelines to remove the data errors and uncertainty. Data uncertainty comes from the inaccurate approximation, so the guidelines in each domain of the research field are set up and gradually improved for enhanced data availability[6,11,23]. First, many existing databases stem from various experimental conditions or measurement techniques or are encoded in different forms (e.g. continuous vectors or tensors, discrete data, weighed graphs), so the standardized databases are necessary for high-quality ML models. Moreover, the majority of databases seldom have permissions for large-scale data extraction access because of the lack of a license and application programming interfaces (APIs)[24]. Accordingly, the homostructure of available data is gradually achieved by generating large-sized, public databases via experimenting or computation.

In terms of nanomaterials data, many public, general-purpose databases have been built based on versatile structures and properties from massive known inorganic molecules and materials. Here, some of the sources of the databases are in the same structures and representatives and also have superior APIs for data access, laying a solid foundation for high-efficiency ML models construction[25]. What's more, the current database provides open-wire calculation platforms to prevent data missing or shortage, consequently for higher-throughput processing capacity, like Fireworks[26] and AiiDA[27].

## 5. Data featurization

Raw data from the databases are usually trained into the form of numerical values, but how data is featured influences the performance of ML models. Ideal featured datasets are useful in guiding the more accurate prediction results, so it is of great value for data preprocessing into appropriate formats, like transforming spectral information into a more readable Fourier frequency domain.

Closely related to the choice of nanomaterials problems, feature engineering should usually be in accordance with the specific structure or composition for efficient one-to-one output in a single molecular/crystal chemical system. The optimum representation indicates the best combinations between the choice of problems and the operation of ML algorithms. Also, overfitting should be prevented via the removal of redundant or highly correlated descriptors. For instance, the lattice constant (containing three angles between the two of x, y and z axis), X-ray diffraction parameters and its crystal structure data of the same specific crystal molecules together comprise the redundant information; the melting and boiling point of one substance inevitably are highly correlated with the cohesive energy. Generally, ML training data, related to the molecular/crystal structure, has a relatively limited correlation with the size of datasets unless the target parameters are of exorbitant value to make the cost of high-throughput calculations negligible[28].

So how are important property parameters of molecules/materials described and converted into feature sets for good input in ML models? Firstly, these datasets originate from the structural characteristics where the positions of atoms and angles between them (serving as the composition-based traits) have been taken into considerations before the general interest in ML. These characteristics, usually containing basic elemental physicochemical properties, are used as descriptors to output other predicted properties[7,29]; however, composition-based featurization only shows satisfactory operating performance with the restricted structural degree of freedom, like aiming at a specific range of crystal prototypes[30].

Next, the second step is to extend the definition range of composition-based structure theoretical system, like considering the features about spatial distance including translation, permutation or rotation of homonuclear atoms, angle, dihedral (HDAD), London Matrix, molecular atomic radial angular distribution (MARAD), and histograms of distance. These distance-based characteristics benefit from the locality of the atomic micro-environment and are further applied to molecular/crystal structures[31]. For example, the Coulomb Matrix requires each atom pair to encode their Coulomb forces for complete information on molecular structures[32]; d-band centre fingerprints of transition-metal compounds, seen as an unclear structure-based feature, are widely exploited to investigate the catalytic activity in hydrogen/oxygen evolution reactions (HER/OER)[33].

Thirdly, graph-based feature engineering gradually gets public concern because one graph can exhibit an engaging picture of each atom and the bonds between atoms. In alkaline/neutral HER, graphs are extensively used to describe the kinetics affected by intermediate spectators and the energetics with the activation barrier formed by intermediates[33,34].

Overall, the choice of feature data in three different representatives is intimately connected with the practical performance and durability of ML models. However, the choice of target problems and the corresponding ML algorithms is determined by domain knowledge, which depends on physicochemical intuitions for appropriate ML feature selection, thereby leading to non-ideal performance and generalizability. Conversely, data-driven data representation can evidently avoid the bias by top-down featurization (down-selecting the feature datasets from a sea of candidate sets of data) such as genetic learners[35]. Particularly, if the selected high-dimensional data is used to generate low-dimension-based features, like the principal component analysis (PCA)[36] for linear relationship and the manifold learning for non-linear relationship[37], so great adjustment in low-dimensional representation will no longer rely highly on the hyperparameters for better predictions in descriptors of electronic band structures[38] and the local Fermi level regulation of perovskite for better photoelectric performance[39], listed here as two typical examples.

## 6. Model selection and optimization

After datasets have been gathered and featurized, the choice of nanomaterials-related learning models is needed for ML model training. The training modes in three categories are supervised, semi-supervised and unsupervised. In supervised learning, a specific function is formed between input features (usually composition- or structure-based characteristics) and output labels/values (usually property parameters), which is relatively mature and currently used most broadly in the field of nanomaterials[18,40]. Due to its low time and money cost, ML-supervised learning has become broad guidelines in the tremendous-speed selection of qualified chemical products and targeted predictions of properties of interest. In semi-supervised learning, only limited sets of feature (labeled) data can be output from excessive amounts of unlabeled data. Unsupervised learning aims to accurately predict the tendencies and emerging patterns by analyzing the unlabeled data more generally, such as the ions and mass transport in electrolytes[41]. In the context of a new-born method, reinforcement learning, instead of specifying how ML models can obtain the mapping from data input, is used to facilitate the performance of models, where rewards or punishments can stimulate the operation of models based on interactions with environments like guiding a time-sequence of predictive Chemical Vapor Deposition (CVD) synthetic parameters[42] and realizing the "on-demand" nucleation of quantum dots with a certain range of emission wavelength and size distribution quality[43]. Then, the standard ML-guided models in nanomaterials design are comprehensively summarized in **Fig. 1**:

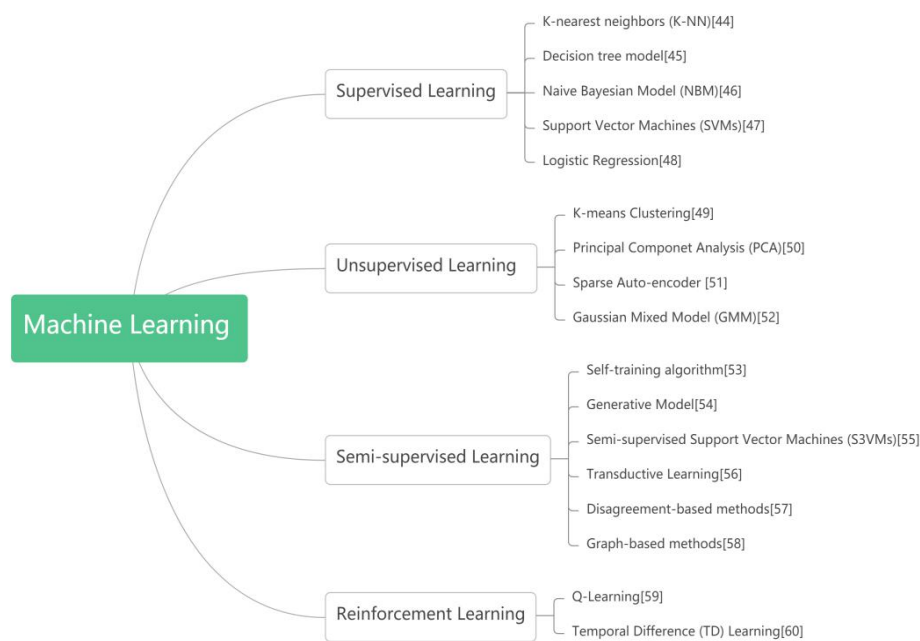

**Fig.1** A classification and the corresponding algorithms of machine learning models in materials design[44-60]

The choice of the model is closely linked with the mapping relationship between input descriptors and output values, where learning repetitively optimizes the performance by adjusting the hyperparameters. It can be inferred that considering model selection and optimization, how experts treat these internal parameters to minimize the loss becomes a crucial issue.

Usually, the losses or prediction errors come from bias, variance and irreducible errors. Specifically, model bias is from mistaken assumptions, which leads to the perplexing understanding or extremely high complexity in the underlying correlation between labeled input and predicted output (underfitting); model variance indicates the sensitivity from the tiny fluctuation of the datasets training, which triggers declined prediction results while the accuracy of data representation increasing (overfitting). Overfitting is common due to the tremendous amount of training data out of the capacity of operating models. Consequently, keeping the balance between the size of input datasets and the simplicity of models is critical[22,61]. Occasionally, more simple and straight-forward models have more potentiality in contributing to the feature revelation of target predictions in new materials design. Ideally, we anticipate a sufficiently low error allowing a relatively clear input-output function (low model bias) and the negligible differences between training and validation errors (low model variance).

However, large numbers of nanomaterials data with high dimensions are easy to make ML models overfitting. Cross-validation (CV) is a common method aiming at size restrictions of datasets, where the data is split into several folds and some of the folds are used as individual data space for averaged model performance via splitting behavior[62]. One drawback is that the method still cannot be effective for many materials models based on entirely different datasets or small-sized sets of feature data. Accordingly, CV cannot be generalized in the whole nanomaterials ML models.

Another set of data also playing an indispensable role in promoting the actual performance of the ML models are hyperparameters. Usually set prior to the data training process, they determine the architecture of input datasets by monitoring the performance of input parameters via adjusting their values properly (usually called pre-training). The altered hyperparameters make the best use of the raw data and optimize the prediction losses and errors for boosted performance[63].

## 7. Application

In the workflow of constructing an ML model, the progress of new nanomaterials design and discovery is further prompted based on the domain knowledge or driving force from available datasets. Although there are increasing numbers of ML tools and packages assisting the development of "intelligent" materials science, like general-purpose tools (e.g. H2O.ai[64], Keras[65], Scikit-learn[66]) and specialized tools for molecular and materials science (e.g. COMBO[67], ANI[68], MEGNet[69]), the critical step is to realize the internal feature mapping between input physical parameters and output values. Here, ML-assisted approaches are used to identify more interpretable descriptors/fingerprints for higher-accuracy, higher-efficiency materials discovery.

Three main categories are highlighted considering the use of ML models in new nanomaterials design and discovery: property predictions, combining theoretical chemistry and new materials discovery from composition- or structure-based features.

## 7.1. Application in property predictions

Instead of experimental measurements or computational simulations, material properties can be more intelligently predicted at a low cost with ML-related techniques. The superiority of ML models is mapping the correlations between properties (labeled or unlabeled) and their related factors (decision attributes) by training models in order to make accurate predictions. Here, versatile material properties can be analyzed at a microscopic or macroscopic level.

Macroscopic predictions primarily focus on the structure-property relationship between mechanical/physical factors and material microarchitectures/microinterfaces[70]. For instance, in terms of rechargeable alkali-ion batteries, the graph-based Crystal Graph Convolutional Neural Networks (CGCNN) model with deep learning algorithms was implemented to graph the ideal interface stabilized with suitable solid electrolytes; here, the proper mechanical stabilization for electrodeposition was realized by fully considering the orientations of electrolytes themselves (isotropy) as well as with metals (anisotropy)[58].

Similarly, microscopic characteristics, including structural/atomic factors, largely determines the macroscopic properties. For example, in a wide range of perovskite-based epitaxial materials, the logistic regression (LR) model is efficient for lattice constant predictions and more apparent relationship with other microscopic parameters like reorganization energy, electronegativity and band gap[48,71]. Further, the prediction accuracy of five different ML models, including LR, kernel ridge regression (KRR), the K-nearest-neighbor (KNN) method, artificial neural network (ANN) and support vector regression (SVR), were comprehensively compared to study the impact of elemental data representation by identifying the minimum of the prediction error of atomization energy[72].

## 7.2. Theoretical-chemistry-combined application

Density-Function-Theory-based (DFT-based) molecular/materials modeling has been a mature tool to contribute to the high-speed analysis of physicochemical properties from structures or compositions. Currently, the main challenges of the prediction speed and accuracy of this class are the increasing complexity of the system and the size of candidate datasets, such as the non-classical approximations on the electron spin densities and the corresponding electron-correlation energy[73]. Based on the unpredictable data availability brought by growing numbers of structure-property datasets, more multi-body descriptors need taking into account. Factually, more feature parameters as input in electron-correlation approximations can possibly realize the clear mapping between self-consistent densities and energy-like aspects, as the example of kinetic energy predictions in non-interacting spinless fermions revealed, proposed by Snyder et al.[13,73,74].

Another class linked with theoretical science is the tempero-spatial analysis of chemical reaction processes, including interfacial dynamics, ion solvation and disordered phases. ML models can optimize the whole process by assessing the system parameters[75]. Especially, in the research field of electrocatalysis, a suitable electrocatalyst requires a well-defined nanosurface with great specific surface area, which makes the surface unsaturated atoms as active sites perfectly combine reactants in electrolytes[33,76].

## 7.3. Application in new materials discovery

The difficulty of ML techniques in design and discovery of nanomaterials is how more accurate and interpretable descriptors can be connected with anticipated properties to identify the previously unknown structure-property relationship. ML models in new materials discovery are currently divided into four parts: structure-oriented design, composition-oriented design, inverse design for desired materials and pharmaceutical design. Different design approaches all obey the aforementioned rules of constructing an ML model, but the discrepancy in their research progresses is the error and the variance generated from the established link between system descriptors and target parameters[77,78].

Structure-oriented methods focus on a particular sort of molecular/crystal structures of unknown materials for stability predictions. However, the data representation of crystal solids cannot be completed in high quality with a sea of datasets as input descriptors[78]. Specifically, independent crystal structure predictions need to take overloaded particulate distribution data into considerations, so a transferable representation between domains can provide more accurate predictions. Racugglia et al. [79] utilized the previously failed reaction data to successfully identify the correlations between crystallization and thermostability of templated vanadium selenites with an SVM-supported model. More importantly, they identified the formation conditions of novel templated products with a prediction error lower than 11%.

Composition-oriented tools are more widely applied to predict the formation of compounds. However, the bottlenecks refer to the limited component-related results based on numerous computational simulations and experiments, thereby hindering the process of new nanomaterials discovery. The general approaches of ML-based composition-oriented investigations can be classified into two categories: (1) introducing various elemental combinations from the elements with the given/known structures; (2) revealing the mechanism of ionic replacement for new nanomaterials. Although structural predictions can occur without prior domains of knowledge, component-related predictions based on the Bayesian model also focus on the collection, mining and processing of prior datasets[80]. The Bayesian model is broadly used in predicting material components due to its extraordinary estimation capabilities on posterior possibility. For example, a high-performance ionic-substitution model based on the stability of crystal structure was proposed to verify the substitution trends [81].

Besides two categories of forward design based on the specific features, inverse design is practised for desired properties or functionalities, but its solutions must be found in many candidate materials, making it tough to analyze[45,82]. Thus, intelligent data-driven experimental screening becomes a general solution to large-scale calculations in design. Here, a certain range of feature data subsets can be used for high-throughput virtual screening to extract information and provide target predictions like environmental safety, nanotoxicology and ecotoxicology of engineered nanomaterials comprehensively detected by Thomas's team [82].

In terms of pharmaceutical design, the construction and analysis of structure-activity correlations are critical for more comprehensive, accurate insight into pharmacological activity. Specifically, one desired parameter of drugs can establish a mapping relationship with the corresponding structural features[83]. For instance, a well-tuned sequence-based generative model (deep reinforcement learning model) for molecular de-novo design was verified to predict the activity against the dopamine receptor type 2 from the desirable structures and generate these ideal biological molecules with over 95% predicted to be active[84].

## 8. Conclusion

Machine Learning has been initially applied in materials science and engineering, exhibiting entirely different perspectives from traditional research modes, including property prediction, new materials discovery and novel combinations with theoretical science, both microscopically and macroscopically. The in-depth research shows more superior advantages of ML-based tools, such as low time consumption, low cost, strong generalization and potentiality of high-throughput calculations. Undoubtedly, the development of ML is revolutionizing the design and discovery of novel nanomaterials by presenting a precise mapping between desired properties and input parameters, thereby successfully tackling the regression, clustering, classification and prior and posterior approximation problems.

However, there are still technical problems to be addressed for more fruitful academic advances. Because of the overfitting generated by excess original data subsets in materials science, prediction accuracy from smaller datasets is needed for more effective learning. In addition, better interpretability is needed to address the "black" box problem; here, more flexible and transferable fingerprints are crucial for the demonstration of underlying physical natures in nanomaterials. Finally, the usability of ML models determines the ability to solve practical, targeted problems in sophisticated systems.

Overall, it's recommended to combine materials science and machine learning for more profitable achievements, where the frontier attaches importance to how high-quality sample data are extracted from a sea of datasets while the latter focuses on how more high-quality data can be amplified from known data for higher-accuracy analysis.